\documentclass[twocolumn]{aastex61}

\begin{document}

\title{A luminous molecular gas pair beyond redshift 7}

\author[0000-0002-0786-7307]{Ekaterina Koptelova}
\affil{Graduate Institute of Astronomy\\
National Central University \\
Taoyuan City, 32001, Taiwan}
\nocollaboration

\author{Chorng-Yuan Hwang}
\affiliation{Graduate Institute of Astronomy\\
National Central University \\
Taoyuan City, 32001, Taiwan}

\nocollaboration







\begin{abstract}

We report the first detection of molecular gas beyond redshift 7.
The molecular gas is associated with the host galaxy of the quasar
candidate PSO J145.5964+19.3565 and its companion PSO
J145.5964+19.3565N separated by 20.7 kpc. The molecular gas of
both companions is detected in two rotational transition lines of
carbon monoxide, CO(6--5) and CO(7--6), with total luminosities of
$L_{\rm CO(6-5)}\approx26\times10^{8}$ and $L_{\rm
CO(7-6)}\approx17\times10^{8}$$L_{\bigodot}$. The two companions
contain (36 -- 54)$\times$10$^{10}$$M_{\bigodot}$ of molecular gas
assuming that their CO spectral line energy distributions are
typical for star-forming galaxies. We also detected the Ly$\alpha$
line of PSO J145.5964+19.3565 at $z\approx7.08$. The Ly$\alpha$
line emission is extended and might represent the blended emission
of two different sources at a separation of $<$5 kpc. The detected
CO and Ly$\alpha$ emission likely originate from a system of
interacting star-forming galaxies that might host a quasar(s). We
also report the detection of a new emission line from the system
which is a possible 793.62-GHz water maser line.

\end{abstract}



\section{Introduction} \label{sec:intro}

Molecular gas may constitute a significant fraction of the total
mass of high-redshift star-forming galaxies and host galaxies of
high-redshift quasars
\citep{2010MNRAS.407.2091G,2013MNRAS.429.3047B,2003Natur.424..406W,2003A&A...409L..47B,2017ApJ...845..154V}.
The physical properties of molecular gas provide important
insights on the formation of massive galaxies and their
coevolution with supermassive black holes.

Carbon monoxide (hereafter CO) is the most common tracer of
molecular gas (composed mainly of molecular hydrogen, H$_{2}$;
\citealp{2005ARA&A..43..677S,2013ARA&A..51..105C}). The current
most distant detections of molecular gas using CO line emission
are from the host galaxies of luminous high-redshift quasars and
massive star-forming galaxies at $6<z<6.9$
\citep{2003Natur.424..406W, 2003A&A...409L..47B,
2010ApJ...714..699W, 2011ApJ...739L..34W, 2011AJ....142..101W,
 2017ApJ...845..154V,2013Natur.496..329R,2017ApJ...842L..15S}. These galaxies are among the largest reservoirs of
 molecular gas known at high redshift.

The total mass and distribution of molecular gas is well-traced by
the CO line emission in ground-state transition CO(1--0)
\citep{1991ARA&A..29..581Y,2005ARA&A..43..677S}. However, this
transition is difficult to observe from high-redshift sources
\citep[e.g.,][]{2006ApJ...650..604R}. At high redshift, the
CO($J$$\sim$4--8) transitions near the peak of the CO spectral
line energy distribution (SLED) of star-forming galaxies become
important tracers of molecular gas
\citep{2009ApJ...703.1338R,2011ApJ...739L..32R,2005A&A...440L..45W,2007A&A...467..955W,2010ApJ...714..699W,
2011ApJ...739L..34W, 2011AJ....142..101W, 2017ApJ...845..154V}.
These CO transitions can potentially be used to search for new
high-redshift star-forming galaxies and quasars, especially for
those which are optically faint, in the obscured phases of their
formation \citep[e.g.,][]{2013ApJ...767...88W,2013Natur.495..344V,
2016ApJ...822...80S, 2017MNRAS.472.2028F}.

In this Letter, we report the discovery of a luminous molecular
gas pair at $z=7.09$ using the CO(6--5) and CO(7--6) line
observations of $z>6.5$ quasar candidates selected from the
Panoramic Survey Telescope and Rapid Response System 1 survey
\citep[PS1;][]{2016arXiv161205560C}. The detected CO emission is
associated with the quasar candidate PSO\,J145.5964+19.3565
(hereafter PSO145+19). The CO redshift of PSO145+19 was confirmed
by the detection of the UV Ly$\alpha$ $\lambda$1216 line at the
similar redshift.

\section{Observations} \label{sec:observations}
\subsection{ALMA observations}

PSO145+19 was observed in Band\,3 of the Atacama Large Millimeter
Array (ALMA) on August 20 and 30, 2016. The observations were
carried out in frequency intervals $\sim$84--92 and 96--104\,GHz
to observe the CO(6--5) and CO(7--6) transitions ($\nu_{\rm
rest}=691.47$ and 806.65\,GHz) from quasar candidates at
$6.5\lesssim z\lesssim7.2$ selected using the following criteria:
$z_{\rm PS1}-y_{\rm PS1}>$2, photometric error $\sigma_{y_{\rm
PS1}}<0.1$, and $z_{\rm PS1}<24$\,AB magnitudes
\citep{2017NatSR...741617K}.

The synthesized beam sizes (spatial resolution) of the data
obtained on August 20 and 30, 2016 were
$\sim0\farcs9\times0\farcs7$ and $0\farcs7\times0\farcs4$,
respectively. The total on-source time at each epoch was about
19\,min. The visibility data were imaged using task CLEAN of the
Common Astronomy Software Applications \citep[CASA
4.7.0;][]{2007ASPC..376..127M} with a cell of 0$\farcs$12, natural
weighting and channel width of 23.44\,MHz ($\sim$80 and
70\,km\,s$^{-1}$ in the lower and upper receiver sidebands,
respectively). We run CLEAN with a taper parameter of 1$\arcsec$
to reduce noise fluctuations. To identify emission lines, we
created intensity maps by sequentially averaging five nearby
frequency channels of data cubes. Each map covered
$\sim$400\,km\,s$^{-1}$, which is the typical widths of the CO
lines detected from the host galaxies of high-redshift quasars
\citep{2010ApJ...714..699W}. The maps were calculated using CASA
task IMMOMENTS with MOMENTS=--1. We checked the maps for emission
peaks with signal-to-noise ratios (SNRs) $>$3 at the expected
frequency range of either CO(6--5) or CO(7--6). We then searched
for the corresponding emission line regardless the SNR of the
latter.

\begin{figure*}[ht!] \vspace{-14cm} \hspace{+100cm}
\epsscale{1.4}\plotone{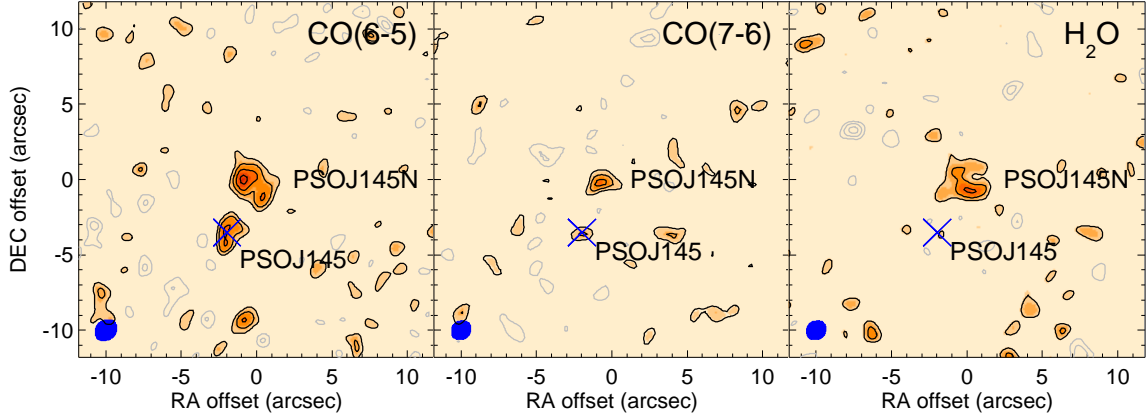} \caption{CO(6--5), CO(7--6) and
H$_{2}$O intensity maps of PSO145+19 and PSO145+19N. The maps were
calculated by averaging the 23.44-MGz channels within velocity
ranges of $\sim$880, 910, and 550\,km\,s$^{-1}$, respectively. The
position of PSO145+19 measured by PS1 is marked with a cross. The
beam size of the CO(6--5) map is $1\farcs4\times1\farcs1$
($7.1\times5.6$\,kpc), and the beam size of the CO(7--6) and
H$_{2}$O maps is $1\farcs2\times1\farcs0$ ($6.1\times5.2$\,kpc)
(ellipses). Contours are drawn at (-3, -2, 2, 3, 4,
5)$\times$$\sigma$, where $\sigma\approx0.2$\,mJy beam$^{-1}$.
Negative contours are shown using gray lines. \label{fig:fig1}}
\end{figure*}

\subsection{Near-infrared observations}

The near-infrared spectrum of PSO145+19 ($4\times950$\,s
exposures) was obtained using the Gemini Multi-Object Spectrograph
\citep[GMOS;][]{2004PASP..116..425H} of the Gemini North Telescope
on October 28, 2017 under seeing of $\sim$0$\farcs$7. We used
grating $R400$ and a slit width of 1$\arcsec$. The spatial
resolution and dispersion of the data were
0$\farcs$1614\,pixel$^{-1}$ and 3.034\,\AA\,pixel$^{-1}$, the
spectral resolution was $R\approx960$. The flux calibration of the
spectrum was performed using the GMOS spectrum of the standard
star EG131 taken at a similar epoch. We also took deep $J$-band
images of PSO145+19 with the UH2.2 telescope of the University of
Hawaii on March 16 and 18, 2017 (see Section~\ref{sec:uvlum}).

\section{Results} \label{sec:results}

\subsection{CO and water emission lines}

\begin{figure*}[ht!]\vspace{-2cm}
\epsscale{1.2}\plotone{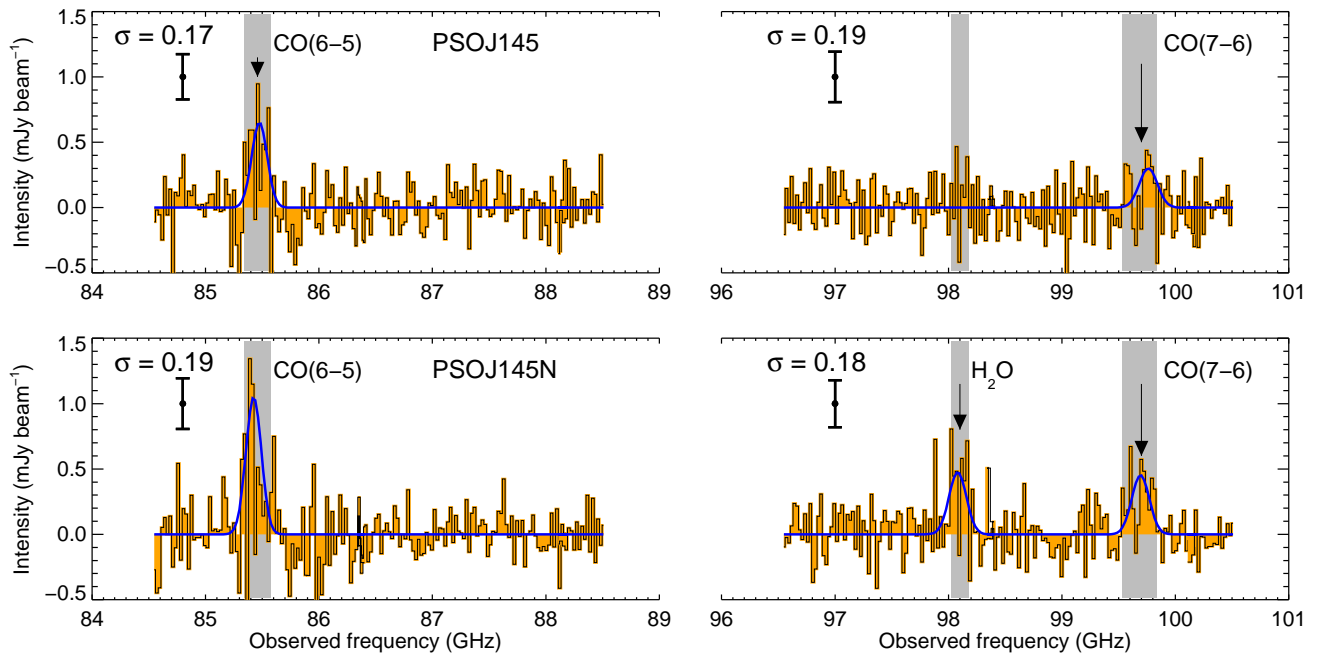} \caption{ALMA spectra of
PSO145+19 and PSO145+19N binned to a resolution of 23.44\,MHz. The
Gaussian profiles of the lines are shown in blue. The shaded
regions mark channels used for the calculation of the intensity
maps. Errorbars represent the typical rms errors.\label{fig:fig2}}
\end{figure*}

In the 400-km\,s$^{-1}$ intensity maps, we identified the CO(6--5)
and CO(7--6) lines at a fixed frequency separation. The emission
in both transitions is spatially extended and shows two luminous
peaks (Figure~\ref{fig:fig1}). The fainter peak coincides with the
PS1 position of PSO145+19, whereas the brighter peak, denoted
PSO145+19N, is located 4$\arcsec$ away. The spectra of PSO145+19
and PSO145+19N presented in Figure~\ref{fig:fig2} were extracted
from the cleaned data cubes by integrating the intensities within
apertures of 4$\arcsec$ in diameters centered on the two peaks.
The aperture size was chosen so as to include more than 90\% of
the emission of PSO145+19N. The observed line profiles were fitted
with Gaussians. The integrated SNRs of the CO(6--5) and CO(7--6)
lines, estimated using spectral channels within the Gaussian
$\pm$2$\sigma$ of the lines, are about 10 and 5 for PSO145+19, and
about 12 and 7 for PSO145+19N. Thus, even the CO(7--6) lines of
PSO145+19 and PSO145+19N are not too significant at the line peaks
(SNR$_{\rm peak}\sim2$ and 3, respectively), the integrated line
SNRs indicate that all detected lines are quite significant and
are not spurious detections. Moreover, the probability of a
spurious detection of two emission lines at a fixed separation is
extremely low. From the normal distribution, the probability of
such a simultaneous event is on the order of 10$^{-29}$. Compared
to these two lines, some other field detections seen in
Figure~\ref{fig:fig1} are not simultaneously detected in the
CO(6--5) and CO(7--6) intensity maps and most likely represent
random peaks.

The velocity-integrated fluxes of the lines, $I_{\rm line}$, were
computed by summing the observed line intensities over spectral
channels within $\pm$2$\sigma$ of the lines as $I_{\rm
line}=N_{\rm beam}$$\Delta\upsilon\Sigma_{i}I_{i}$, where $I_{i}$
is the line intensity in channel $i$ in Jy\,beam$^{-1}$,
$\Delta\upsilon$ is the velocity width of the channel in
km\,s$^{-1}$, and $N_{\rm beam}$ is the number of synthesized
beams in the aperture. The flux uncertainties were calculated as
$\sigma_{I_{\rm line}}=N_{\rm
beam}$$\Delta\upsilon$$\sqrt{\Sigma_{i}\sigma_{i}^{2}}$, where
$\sigma_{i}$ is the rms error in channel $i$ in mJy\,beam$^{-1}$.
The luminosities of the emission lines were computed using
formulae (1) and (3) of \citet{2005ARA&A..43..677S}.

The redshifts of PSO145+19 and PSO145+19N, measured as the mean
redshifts of the CO(6--5) and CO(7--6) transitions, are similar
within the uncertainties. The widths of the CO(6--5) and CO(7--6)
lines were calculated as the Gaussian full width at half maximum
(FWHM) (see Table~\ref{tab:table1}). The linear distance between
PSO145+19 and PSO145+19N is 20.7\,kpc (assuming a cosmology with
$H_{\rm 0}=70$\,km\,s$^{-1}$\,Mpc$^{-1}$, $\it \Omega_{\rm
m}=\rm0.3$, and $\it \Omega_{\rm \Lambda}=\rm0.7$). The emission
from the dust continuum of PSO145+19 and PSO145+19N was not
detected. The three-sigma upper limit on the flux of the dust
continuum estimated at the position of PSO145+19 is 0.99\,mJy.

We also detected a third strong emission line in the spectrum of
PSO145+19N with a line SNR of 5. This line is identified as a
possible H$_{2}$O($J_{K_{a}K_{c}}$=1$_{10}$--1$_{01}$) water maser
line at $\nu_{\rm rest}=793.62$\,GHz predicted by the calculations
of \citet{2016MNRAS.456..374G}. According to
\citet{2016MNRAS.456..374G}, the 793.62-GHz water emission
originates from the dense and hot gas with densities of $n_{\rm
H_{2}}>10^{6}$\,cm$^{-3}$ and kinetic temperatures of $T_{\rm
kin}\gtrsim1000$\,K. Compared to this water line, the CO emission
traces less dense and colder gas with a density and temperature of
$n_{\rm H_{2}}\sim10^{3}$--10$^{5}$\,cm$^{-3}$ and $T_{\rm
kin}\sim100$\,K \citep{2009ApJ...703.1338R, 2011ApJ...739L..32R,
2005A&A...440L..45W, 2007A&A...467..955W}.

To estimate the spatial extent of the CO emission of PSO145+19 and
PSO145+19N, we fitted their CO(6--5) intensity distributions with
two-dimensional Gaussians using CASA's task IMFIT. The
characteristic sizes of these two CO companions were estimated as
a quadratic mean of their FWHMs along the major and minor axes. We
summarized the measured properties of the CO emission of PSO145+19
and PSO145+19N in Table~\ref{tab:table1}.

\begin{deluxetable*}{l|cc}
\tablecaption{Properties of the CO and Ly$\alpha$ emission of
PSO145+19 and PSO145+19N\label{tab:table1}} \tablehead{
\colhead{Parameter} & \colhead{PSO J145+19} & \colhead{PSO J145+19N} \\
} \startdata
RA (hh:mm:ss.ss) & 09:42:23.14 & 09:42:23.00 \\
DEC (dd:mm:ss.ss) & 19:21:23.55 & 19:21:27.02 \\
$z_{\rm CO}$ & 7.092$\pm$0.006 & 7.093$\pm$0.004 \\
$z_{\rm H_{2}O}$  &                & 7.092$\pm$0.004 \\
FWHM$_{\rm CO(6-5)}$ (km s$^{-1}$) & 565$\pm$90 & 589$\pm$79 \\
FWHM$_{\rm CO(7-6)}$ (km s$^{-1}$) & 565$\pm$120 & 547$\pm$98 \\
FWHM$_{\rm H_{2}O}$ (km s$^{-1}$) &             & 551$\pm$86 \\
$I_{\rm CO(6-5)}$ (Jy km s$^{-1}$) & 2.81$\pm$0.36 & 3.12$\pm$0.40 \\
$I_{\rm CO(7-6)}$ (Jy km s$^{-1}$) & 1.52$\pm$0.38 & 1.80$\pm$0.35 \\
$I_{\rm H_{2}O}$ (Jy km s$^{-1}$) &                & 1.96$\pm$0.36 \\
$L_{\rm CO(6-5)}$ (10$^{8}L_{\bigodot}$) & 12.3$\pm$1.6 & 13.6$\pm$1.7 \\
$L^{'}_{\rm CO(6-5)}$ (10$^{10}$ K km s$^{-1}$ pc$^{2}$) & 11.6$\pm$1.5 & 12.9$\pm$1.7 \\
$L_{\rm CO(7-6)}$ (10$^{8}L_{\bigodot}$) & 7.7$\pm$1.9 & 9.2$\pm$1.8 \\
$L^{'}_{\rm CO(7-6)}$ (10$^{10}$ K km s$^{-1}$ pc$^{2}$) & 4.6$\pm$1.2 & 5.5$\pm$1.1 \\
$L_{\rm H_{2}O}$ (10$^{8}L_{\bigodot}$) &       & 9.8$\pm$1.8 \\
$L^{'}_{\rm H_{2}O}$ (10$^{10}$ K km s$^{-1}$ pc$^{2}$) &   & 6.1$\pm$1.1 \\
$M_{\rm H_{2}}$\tablenotemark{a} (10$^{10}M_{\bigodot}$) & 25.7$\pm$2.9 & 29.1$\pm$3.1 \\
$M_{\rm H_{2}}$\tablenotemark{b} (10$^{10}M_{\bigodot}$) & 16.2$\pm$1.9 & 17.8$\pm$1.9 \\
Size (kpc) & 8.2$\pm$3.2 & 12.7$\pm$1.6 \\
$M_{\rm dyn}$ (10$^{10}M_{\bigodot}$) & 36.6$\pm$13.8 &  61.7$\pm$7.8 \\
$z_{\rm Ly\alpha,1}$ & 7.083$\pm$0.003 &   \\
$z_{\rm Ly\alpha,2}$ & 7.086$\pm$0.003 &   \\
FWHM$_{\rm Ly\alpha,1+2}$ (km s$^{-1}$) & 539$\pm$36 &   \\
$L_{\rm Ly\alpha,1+2}$ ($10^{43}$ erg s$^{-1}$)& 5.28$\pm$0.22 &   \\
\enddata
\tablenotetext{a}{Assuming the CO SLED of SMM J16359+6612.}
\tablenotetext{b}{Assuming the CO SLED of IRAS F10214+4724.}
\end{deluxetable*}

\subsection{Ly$\alpha$ line} \label{sec:uvlum}

\begin{figure}[ht!]
\epsscale{1.3}\plotone{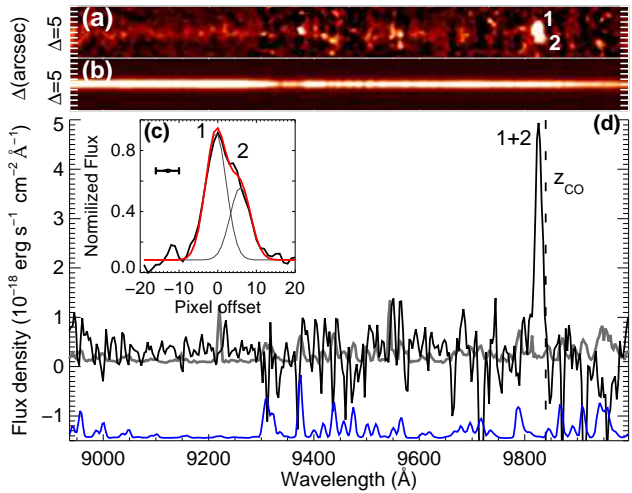} \caption{Ly$\alpha$ line emission
of PSO145+19. The two-dimensional spectrum of PSO145+19 is shown
in comparison with the spectrum of the field star (panels
\textbf{a} and \textbf{b}). The observed spatial profile of the
Ly$\alpha$ emission along the most extended direction is shown in
panel \textbf{c} with a thick black line. The two Gaussians fitted
into the spatial profile and their sum are shown with gray and red
lines. The FWHM size of the field star is shown with an errorbar.
Panel \textbf{d} shows the total spectrum of 1 and 2. The error
spectrum is shown by a thick gray line. The sky emission lines are
shown in blue. The wavelength corresponding to the redshift of the
CO emission of PSO145+19 is marked with a dashed line.
\label{fig:fig3}}
\end{figure}

The Ly$\alpha$ emission line of PSO145+19 was detected at the
wavelength interval which is not affected by sky emission lines.
The line appears to be about two times more extended in the
spatial direction compared to the spatial profile of a field star
within 17$\arcsec$ of PSO145+19 observed simultaneously in the
same slit. Thus, PSO145+19 might represent the blend of two
Ly$\alpha$ lines from two different sources separated by less than
1$\arcsec$ (Figure~\ref{fig:fig3}a). At this separation the
spectra of the sources are not fully resolved and their
overlapping Ly$\alpha$ lines appear extended in the spatial
direction. As shown in Figure~\ref{fig:fig3}c, the spatial profile
of the Ly$\alpha$ emission along its most extended direction can
be described well by two blended Gaussians of the same widths
equal to the FWHM of the field star.

Further, we denote the two Ly$\alpha$ peaks as sources 1 and 2.
The total spectrum (extracted using an aperture of 12 pixels),
which includes the Ly$\alpha$ emission of sources 1 and 2, is
presented in Figure~\ref{fig:fig3}d. The FWHM and luminosity of
the Ly$\alpha$ line estimated from the Gaussian fit of the total
spectrum are given in Table~\ref{tab:table1}. The individual
spectra of sources 1 and 2 were extracted using two adjacent
apertures of 6 pixels in widths centered on each source. The
redshifts of these two sources estimated from the Gaussian fit of
their Ly$\alpha$ lines, are $z=7.083\pm0.003$ and $7.086\pm0.003$
with a small velocity offset of $\sim$120\,km\,s$^{-1}$. The
approximate FWHMs of the Ly$\alpha$ lines of 1 and 2 are $\sim$490
and 630\,km\,s$^{-1}$. The Ly$\alpha$ lines of sources 1 and 2 are
blueshifted with respect to the CO lines by about 340 and
210\,km\,s$^{-1}$. Similar blueshifts of 100--3000\,km\,s$^{-1}$
of the UV emission lines relative to molecular and atomic lines
have been observed in high-redshifts quasars
\citep{2018ApJ...856L...5F,2016ApJ...816...37V,2017ApJ...849...91M,koptelova2019}.
The nearby UV continuum around the Ly$\alpha$ line emission was
not detected and therefore, its estimated luminosity should be
considered as a lower limit. The weak continuum is visible between
sky emission residuals at wavelengths $<$9300~\AA. The mean flux
density of the continuum in region 9000--9300\,\AA, not
contaminated by strong sky emission lines, is
($0.3\pm0.2$)$\times$10$^{-18}$\,erg\,s$^{-1}$\,cm$^{-2}$\,A$^{-1}$
and its average SNR is $\sim$2. The detection of the UV continuum
blueward of the Ly$\alpha$ line might indicate inhomogeneous
reionization at redshift around 7
\citep[e.g.,][]{2015MNRAS.447.3402B,2017A&A...601A..16B}.

The $J$-band brightness of PSO145+19 measured from the UH2.2
images is $m(J)=21.87\pm0.17$\,AB. There was no $J$-band source
detected at the position of PSO145+19N. If PSO145+19N has a
counterpart in the $J$ band, it should be fainter than a 5-sigma
limiting magnitude of 22.4\,AB. Such faint sources are below the
5-sigma detection limit of PS1 in the $y_{\rm PS1}$ band (21\,AB).
To select PSO145+19, we used the first and second internal
releases of the PS1 catalogs \citep[see][]{2017NatSR...741617K}.
In these catalogs, the brightness of PSO145+19 measured from stack
images is $m(y_{\rm PS1})=19.40\pm0.06$. However, we note that
PSO145+19 was not detected by PS1 at almost all epochs except for
one when its brightness was $m(y_{\rm
PS1})=18.64\pm0.02$\,AB\footnote{https://catalogs.mast.stsci.edu/}
(December 8, 2011). Only due to the contribution of this
particular epoch into the $y$-band stack flux of PSO145+19, it was
selected by us as a high-redshift quasar candidate. One month
later, PSO145+19 was also detected in the $z_{\rm PS1}$ band with
$m(z_{\rm PS1})=18.72\pm0.04$\,AB. The $z_{\rm PS1}$-/$y_{\rm
PS1}$-band brightness of PSO145+19 at the epoch of our Gemini/GMOS
observations, constrained from its total spectrum
(Figure~\ref{fig:fig3}d), is $\gtrsim$24\,AB magnitudes.

\section{Discussion}

The detected CO(6--5) and CO(7--6) line emission from the quasar
candidate PSO145+19 revealed that it is associated with two
luminous and extended molecular gas companions, PSO145+19 and
PSO145+19N. Similar CO-luminous binary companions exhibit the host
galaxies of some high-redshift quasars
\citep[e.g.,][]{2012A&A...545A..57S,2004ApJ...615L..17W}.

The spatially extended Ly$\alpha$ line of PSO145+19 further
suggests that it might represent a system of two sources at
slightly different redshifts separated by $<$5\,kpc
($<$1$\arcsec$). These two Ly$\alpha$ sources are not resolved in
the CO line emission. The widths of the Ly$\alpha$ lines of
PSO145+19 1 and 2 are consistent with those of some high-redshift
star-forming galaxies and quasars
\citep[e.g.,][]{2018PASJ...70S..35M}. The Ly$\alpha$ luminosity of
PSO145+19, that is somewhat higher than the typical Ly$\alpha$
luminosities of star-forming galaxies
($\lesssim$10$^{43}$\,erg\,s$^{-1}$)\citep{2018PASJ...70S..35M},
and evidence for variability of this source favor its
interpretation as a quasar (or a quasar pair). The brightening of
PSO145+19 detected in the $z_{\rm PS1}$ and $y_{\rm PS1}$ bands
might be caused by significant variability of its continuum.
Assuming that this variation is driven by a UV flare, we estimate
an accretion disk size of log$(R/$cm$)=$log$(ct)=16.6$, where $t$
is the PS1 cadence of 4 months in the rest frame of PSO145+19
adopted as the upper limit on the timescale of the observed
variation. The inferred size is consistent with the UV sizes of
the accretion disks of quasars estimated using microlensing
\citep{2011ApJ...729...34B}. Thus, the two luminous CO companions
(PSO145+19 and PSO145+19N) and the Ly$\alpha$ emission of
PSO145+19 at the similar redshifts indicate that PSO145+19,
initially selected as a single source quasar candidate, might
represent a system of several interacting galaxies possibly
hosting a quasar(s). The CO luminosity of the system indicates
intensive star formation that might have been triggered by the
interaction between the galaxies. The previous observations of the
host galaxies of high-redshift quasars and submillimeter galaxies
suggest that their star formation is most intensive in very
compact central regions of $\lesssim$1\,kpc
\citep[e.g.,][]{2009Natur.457..699W,2013Natur.496..329R,2015ApJ...810..133I,2017ApJ...837..146V}.
The water line emission is usually associated with these central
regions \citep{2011ApJ...741L..37B,2013Natur.496..329R}. The water
emission of PSO145+19N might also indicate intensive star
formation within the dense gas of PSO145+19N, but on larger
spatial scales than typically (greater than a beam size of
$\sim$5\,kpc). This star formation could be happening in several
interacting (proto-)galaxies unresolved in our observations, or in
the unresolved star-forming cores within a massive gas-rich
galaxy.

\subsection{CO excitation and molecular gas mass}

The relative fluxes of the CO(6--5) and CO(7--6) lines of
PSO145+19 and PSO145+19N ($I_{\rm CO(7-6)}/I_{\rm CO(6-5)}
\thickapprox0.54$ and 0.58, respectively) indicate that their CO
SLEDs peak at transitions $J\lesssim6$. As the flux ratios of
PSO145+19 and PSO145+19N are similar, their CO SLEDs are also
likely similar. However, compared to the host galaxies of quasars
and star-forming galaxies known at $z\lesssim7$, the molecular gas
of PSO145+19 and PSO145+19N appears to be somewhat less excited.
The flux ratios between the CO(6--5) and CO(7--6) lines of these
$z\lesssim7$ galaxies are typically $I_{\rm CO(7-6)}/I_{\rm
CO(6-5)}\gtrsim1$
\citep{2009ApJ...703.1338R,2017ApJ...845..154V,2013Natur.496..329R,2017ApJ...842L..15S}.
The estimated CO(6--5) and CO(7--6) luminosities of PSO145+19 and
PSO145+19N are a few times higher than those of optically luminous
$z>6$ quasars
\citep{2009ApJ...703.1338R,2010ApJ...714..699W,2017ApJ...845..154V},
but comparable to the luminosities of some massive star-forming
galaxies at $z>6$ \citep{2013Natur.496..329R,2017ApJ...842L..15S}.

The mass of molecular gas is related to CO luminosity as $M_{\rm
H2}=\alpha$$L$$^{'}_{\rm CO(1-0)}$, where $\alpha$ is the
conversion factor \citep{2005ARA&A..43..677S}. We estimated the
CO(1--0) luminosities of PSO145+19 and PSO145+19N using the CO
SLEDs of known sources with the relative fluxes between
transitions CO(7--6) and CO(6--5) similar to those of PSO145+19
and PSO145+19N. Among the well-studied star-forming galaxies, the
excitation conditions of molecular gas of PSO145+19 and PSO145+19N
might be similar to those of the submillimeter galaxy SMM
J16359+6612 at $z=2.527$ with $I_{\rm CO(7-6)}$/$I_{\rm
CO(6-5)}=0.49$ \citep{2005A&A...434..819K,2005A&A...440L..45W} and
the quasar IRAS F10214+4724 at $z=2.286$ with $I_{\rm
CO(7-6)}$/$I_{\rm CO(6-5)}=0.77$
\citep{2008A&A...491..747A,2011ApJ...739L..32R}. For SMM
J16359+6612, the luminosity ratios between the CO(6--5), CO(7--6)
and CO(1--0) transitions are $L^{'}$$_{\rm CO(6-5)}$/$L^{'}$$_{\rm
CO(1-0)}\approx0.37$ and $L^{'}$$_{\rm CO(7-6)}$/$L^{'}$$_{\rm
CO(1-0)}\approx0.13$ \citep{2005A&A...440L..45W}. For IRAS
F10214+4724, these ratios are $L^{'}$$_{\rm
CO(6-5)}$/$L^{'}$$_{\rm CO(1-0)}\approx0.52$ and $L^{'}$$_{\rm
CO(7-6)}$/$L^{'}$$_{\rm CO(1-0)}\approx0.29$. Using these
relations, the CO(1--0) luminosities of PSO145+19 and PSO145+19N
were estimated as the weighted means of the CO(1--0) luminosities
independently derived from the CO(6--5) and CO(7--6) lines.

The CO(1--0) flux of high-redshift galaxies is usually converted
into the mass of molecular gas using a conversion factor of $\it
\alpha=\rm0.8$$M_{\odot}$\,(K\,km\,s$^{-1}$\,pc$^{2}$)$^{-1}$
inferred from the observations of nearby star-forming galaxies
\citep{2005ARA&A..43..677S}. Assuming that the gas-to-mass
conversion factor is similar for low- and high-redshift
star-forming galaxies, the estimated CO(1--0) luminosities of
PSO145+19 and PSO145+19N imply that they contain on the order of
10$^{11}$$M_{\bigodot}$ of molecular hydrogen
(Table~\ref{tab:table1}). The gas masses of PSO145+19 and
PSO145+19N derived using the CO SLEDs of SMM J16359+6612 and IRAS
F10214+4724 are comparable. The differences in the gas masses
using these two comparison SLEDs reflect the uncertainties related
to the unknown shape of the CO SLEDs of PSO145+19 and PSO145+19N.
However, both estimates indicate the large masses of the molecular
gas of PSO145+19 and PSO145+19N and place them among the largest
reservoirs of molecular gas currently known at high redshift. For
comparison, the previously reported gas masses of the host
galaxies of high-redshift quasars and massive star-forming
galaxies are typically on the order of
10$^{10}-10^{11}$$M_{\bigodot}$
\citep{2010ApJ...714..699W,2017ApJ...845..154V,2013Natur.496..329R,2017ApJ...842L..15S}.

\subsection{Spatial extent and dynamical mass}

The characteristic sizes (radii) of the CO(6--5) emission of
PSO145+19 and PSO145+19N are $\sim$4 and 6\,kpc. We estimated the
dynamical masses of PSO145+19 and PSO145+19N using the isotropic
virial estimator with $M_{\rm dyn}=2.8\times10^{5}$(FWHM)$^{2}R$,
where FWHM is the CO(6--5) line widths and $R$ is the radius of
the CO(6--5) emission in kpc
\citep{2008gady.book.....B,2013MNRAS.429.3047B,2013Natur.496..329R}
(see Table~\ref{tab:table1}). Thus, molecular gas might constitute
more than 45\% and 30\% of the dynamical masses of PSO145+19 and
PSO145+19N, respectively.

The discovered system might be of particular interest for studying
physical conditions of star-forming molecular gas at early
evolutionary stages of galaxy formation. Further, deeper
observations in the near-infrared/submm wavelengths are needed to
strengthen the results reported here.

\acknowledgments

This work was supported by the Ministry of Science and Technology
of Taiwan, grant Nos MOST105-2119-M-007-022-MY3, and
MOST107-2119-M-008-009-MY3. Based on data obtained with ALMA
(Program 2015.1.01452.S). ALMA is a partnership of ESO
(representing its member states), NSF (USA) and NINS (Japan),
together with NRC (Canada), MOST and ASIAA (Taiwan), and KASI
(Republic of Korea), in cooperation with the Republic of Chile.
The Joint ALMA Observatory is operated by ESO, AUI/NRAO and NAOJ.
Based on observations (Program GN-2017B-DD-4) obtained at the
Gemini Observatory, which is operated by the Association of
Universities for Research in Astronomy, Inc., under a cooperative
agreement with the NSF on behalf of the Gemini partnership: the
National Science Foundation (United States), the National Research
Council (Canada), CONICYT (Chile), Ministerio de Ciencia,
Tecnolog{\'i}a e Innovaci{\'o} Productiva (Argentina), and
Minist{\'e}rio da Ciencia, Tecnologia e Inovac{\~a}o (Brazil).


\begin{thebibliography}{}

\bibitem[Ao et al. (2008)]{2008A&A...491..747A} Ao, Y., Wei{\ss}, A., Downes, D., et al. \ 2008,
\aap, 491, 747

\bibitem[Barnett et al.(2017)]{2017A&A...601A..16B} Barnett, R., Warren, S.~J., Becker, G.~D, et al.\ 2017, \aap, 601,
A16

\bibitem[Becker et al.(2015)]{2015MNRAS.447.3402B} Becker, G.~D., Bolton, J.~S., Madau, P., et al.\ 2015, \mnras, 447,
3402

\bibitem[Bertoldi et al.(2003)]{2003A&A...409L..47B} Bertoldi, F., Cox, P., Neri, R., et al.\ 2003, \aap, 409,
L47

\bibitem[Binney \& Tremaine (2008)]{2008gady.book.....B} Binney, J. \& Tremaine, S.\ 2008, Galactic Dynamics (2nd ed.; Princeton, NJ:
Princeton Univ. Press)

\bibitem[Blackburne et al. (2011)]{2011ApJ...729...34B} Blackburne, J.~A., Pooley, D., Rappaport, S., et al. \ 2011, \apj, 729,
34

\bibitem[Bothwell et al. (2013)]{2013MNRAS.429.3047B} Bothwell, M.~S., Smail, I., Chapman, S.~C., et al. \ 2013, \mnras, 429,
3047

\bibitem[Bradford et al. (2011)]{2011ApJ...741L..37B} Bradford, C.~M., Bolatto, A.~D., Maloney, P.~R., et al. \ 2011, \apjl, 741,
L37

\bibitem[Carilli \& Walter (2013)]{2013ARA&A..51..105C} Carilli, C.~L. \& Walter, F.\ 2013, \araa, 51,
105
\bibitem[Chambers et al. (2016)]{2016arXiv161205560C} Chambers, K.~C., Magnier, E.~A., Metcalfe, N., et al. \ 2016, eprint arXiv:1612.05560

\bibitem[Fan et al. (2018)]{2018ApJ...856L...5F} Fan, L., Knudsen, K.~K., Fogasy, J., et al. \ 2018, \apj, 856, L5
\bibitem[Fudamoto et al. (2017)]{2017MNRAS.472.2028F} Fudamoto, Y., Ivison, R.~J., Oteo, I., et al. \ 2017, \mnras, 472,
2028

\bibitem[Genzel et al. (2010)]{2010MNRAS.407.2091G} Genzel, R., Tacconi, L.~J., Gracia-Carpio, J., et al. \ 2010, \mnras, 407,
2091
\bibitem[Gray et al. (2016)]{2016MNRAS.456..374G} Gray, M.~D., Baudry, A., Richards, A.~M.~S., et al. \ 2016, \mnras, 456, 374
\bibitem[Hook et al. (2004)]{2004PASP..116..425H} Hook, I.~M., J{\o}rgensen, I., Allington-Smith, J.~R., et al. \ 2004, \pasp, 116, 425

\bibitem[Ikarashi et al. (2015)]{2015ApJ...810..133I} Ikarashi, S., Ivison, R.~J., Caputi, K.~I., et al. \ 2015, \apj, 810,
133

\bibitem[Kneib et al. (2005)]{2005A&A...434..819K} Kneib, J.-P., Neri, R., Smail, I., et al. \ 2005,
\aap, 434, 819
\bibitem[Koptelova et al. (2017)]{2017NatSR...741617K} Koptelova, E., Hwang, C.-Y., Yu, P.-C., et al. \ 2017, Scientific Reports, 7, 41617
\bibitem[Koptelova et al. (2019)]{koptelova2019} Koptelova, E., Hwang, C.-Y., Malkan, M.~A., et al. \ 2019, \apj,
in press

\bibitem[Matsuoka et al. (2018)]{2018PASJ...70S..35M} Matsuoka, Y., Onoue, M., Kashikawa, N. et al. \ 2018,
\pasj, 70, S35

\bibitem[Mazzucchelli et al. (2017)]{2017ApJ...849...91M} Mazzucchelli, C., Ba{\~n}ados, E., Venemans, B.~P., et al. \ 2017, \apj, 849,
91
\bibitem[McMullin et al. (2007)]{2007ASPC..376..127M} McMullin, J.~P., Waters, B., Schiebel, D., et al. \ 2007, \pasp, 376, 127
\bibitem[Riechers et al. (2006)]{2006ApJ...650..604R} Riechers, D.~A., Walter, F., Carilli, C.~L., et al. \ 2006, \apj, 650, 604
\bibitem[Riechers et al. (2009)]{2009ApJ...703.1338R} Riechers, D.~A., Walter, F., Bertoldi, F., et al. \ 2009, \apj, 703, 1338
\bibitem[Riechers et al. (2011)]{2011ApJ...739L..32R} Riechers, D.~A., Carilli, C.~L., Maddalena,
R.~J., et al. \ 2011, \apj, 739, L32

\bibitem[Riechers et al. (2013)]{2013Natur.496..329R} Riechers, D.~A., Bradford, C.~M., Clements, D.~L., et al. \ 2013, \nat, 496, 329

\bibitem[Salom{\'e} et al. (2012)]{2012A&A...545A..57S} Salom{\'e}, P., Gu{\'e}lin, M., Downes, D., et al. \ 2012, \aap, 545,
A57
\bibitem[Solomon \& Vanden Bout (2005)]{2005ARA&A..43..677S} Solomon, P.~M., \& Vanden Bout, P.A.\ 2005, \araa, 43,
677
\bibitem[Strandet et al. (2016)]{2016ApJ...822...80S} Strandet, M.~L., Weiss, A., Vieira, J.~D., et al. \ 2016, \apj, 822,
80

\bibitem[Strandet et al. (2017)]{2017ApJ...842L..15S} Strandet, M.~L., Weiss, A., De Breuck, C., et al. \ 2017, \apjl, 842,
L15

\bibitem[Venemans et al. (2016)]{2016ApJ...816...37V} Venemans, B.~P., Walter, F., Zschaechner, L., et al. \ 2016,
\apj, 816, 37
\bibitem[Venemans et al.(2017a)]{2017ApJ...845..154V} Venemans, B.~P., Walter, F., Decarli, R., et al.\ 2017, \apj, 845,
154
\bibitem[Venemans et al. (2017b)]{2017ApJ...837..146V} Venemans, B.~P., Walter, F., Decarli, R., et al.\ 2017, \apj, 837,
146

\bibitem[Vieira et al. (2013)]{2013Natur.495..344V} Vieira, J.~D., Marrone, D.~P., Chapman, S.~C., et al.\ 2013, \nat, 495,
344

\bibitem[Walter et al.(2003)]{2003Natur.424..406W} Walter, F., Bertoldi, F., Carilli, C., et al.\ 2003, \nat, 424,
406
\bibitem[Walter et al. (2004)]{2004ApJ...615L..17W} Walter, F., Carilli, C., Bertoldi, F., et al. \ 2004, \apj, 615, L17

\bibitem[Walter et al. (2009)]{2009Natur.457..699W} Walter, F., Riechers, D., Cox, P., et al. \ 2009, \nat, 457,
699

\bibitem[Wang et al.(2010)]{2010ApJ...714..699W} Wang, R, Carilli, C.L., Neri, R., et al.\ 2010, \apj, 714, 699
\bibitem[Wang et al.(2011a)]{2011ApJ...739L..34W} Wang, R., Wagg, J., Carilli, C.~L., et al.\ 2011, \apjl, 739,
L34
\bibitem[Wang et al.(2011b)]{2011AJ....142..101W} Wang, R., Wagg, J., Carilli, C.~L., et al.\ 2011, \aj, 142,
101
\bibitem[Wei{\ss} et al. (2005)]{2005A&A...440L..45W} Wei{\ss}, A., Downes, D., Walter, F., et al. \ 2005, \aap, 440,
L45
\bibitem[Wei{\ss} et al. (2007)]{2007A&A...467..955W} Wei{\ss}, A., Downes, D., Neri, R., et al. \ 2007, \aap, 467,
955
\bibitem[Wei{\ss} et al. (2013)]{2013ApJ...767...88W} Wei{\ss}, A., De Breuck, C., Marrone, D.~P., et al. \ 2013, \apj, 767,
88

\bibitem[Young \& Scoville (1991)]{1991ARA&A..29..581Y} Young, J.~S. \& Scoville, N.~Z.\ 1991, \araa, 29,
581


\end{thebibliography}
\end{document}